\documentclass[aps,prl,twocolumn,groupedaddress,showpacs]{revtex4}

\usepackage{graphicx}
\begin{document}

\title{Origin of Hysteresis in a Proximity Josephson Junction}

\author{H. Courtois$^{1,2}$, M. Meschke$^{1}$, J. T. Peltonen$^{1}$ and J. P. Pekola$^{1}$}
\affiliation{$^{1}$Low Temperature Laboratory, Helsinki University of Technology, P.O. Box 3500, 02015 TKK, Finland\\
$^{2}$Institut N\' eel, CNRS and Universit\'e Joseph Fourier, 25 Avenue des Martyrs, BP 166, 38042 Grenoble, France}


\date{\today}

\begin{abstract}
We investigate hysteresis in the transport properties of Superconductor - Normal metal - Superconductor (S-N-S) junctions at low temperatures by measuring directly the electron temperature in the normal metal. Our results demonstrate unambiguously that the hysteresis results from an increase of the normal metal electron temperature once the junction switches to the resistive state. In our geometry, the electron temperature increase is governed by the thermal resistance of the superconducting electrodes of the junction.
\end{abstract}


\maketitle

Dissipationless supercurrent can flow between two superconductors, up to a critical current $I_c$ in a Josephson junction. The junction dynamics can be described by the Resistively and Capacitively Shunted Junction (RCSJ) model \cite{Likharev}. The junction capacitance is then responsible for hysteresis in the current-voltage characteristic. In lateral junctions, the distance between the two superconducting electrodes induces an extremely small capacitance, much lower than those in a typical tunnel junction. A non-hysteretic (overdamped) current-voltage characteristic is then expected. Nevertheless, a significant hysteresis is routinely observed in lateral junctions as soon as their critical current is large: once the junction has switched to the resistive branch, it does not recover the superconducting state until the bias current is decreased to a significantly smaller retrapping current $I_r$. This observation does not depend on the nature of the weak link, as it was early observed in superconducting constrictions and microbridges \cite{JAP-Fulton, JAP-Skocpol,JAP-Song}, and more recently in superconducting nanowires \cite{PRL-Bezryadin}, normal metals \cite{PRB-Angers,PRB-Birge}, two-dimensional electron gases (2-DEG) \cite{PRB-Krasnov}, semiconductor nanowires \cite{Science-DeFranceschi}, carbon nanotubes \cite{NatureNanotech-WW} and graphene \cite{Nature-Morpurgo}. Two main explanations have been proposed. Firstly, the Joule power deposited in the weak link can induce a self-heating process so that the local temperature in the normal part increases \cite{JAP-Skocpol, JAP-Fulton, PRB-Tinkham}. Secondly, it has been proposed that the response time of the junction $R_nC$ should be replaced by a time $\hbar/\Delta$ related to the superconducting gap $\Delta$ \cite{JAP-Song} or by the electron diffusion time $L^2/D$ through the junction \cite{PRB-Angers}. Here $R_n$ is the normal state resistance, $L$ is the junction length and $D$ is the electron diffusion constant. These latter approaches are equivalent to assuming an effective capacitance larger than the geometric capacitance. Despite relatively good agreement with the experiments, no strong justification can be brought to support these hypotheses.

In this Letter, we report an experimental study of the hysteretic regime of proximity Superconductor - Normal metal - Superconductor (S-N-S) Josephson junctions. Our results demonstrate unambiguously that the hysteresis results from the increase of the normal metal electron temperature once the junction switches to the resistive state. An electron temperature of up to 0.6 K is measured while the thermal bath remains at 50 mK. We show that, in our geometry, the electron temperature increase is governed by the thermal resistance of the superconducting electrodes.

\begin{figure}[t]
\begin{center}
\includegraphics[width=8.5 cm]{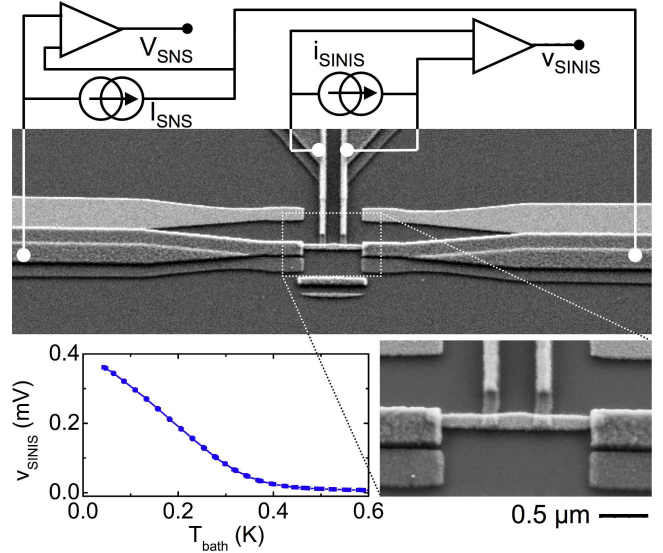}
\caption{(Color online)  Top: SEM image of sample 2 containing a S-N-S junction of 1.5 $\mu$m length with a sketch of the measurement circuit. Two tunnel probes (top of the image) are connected to the normal metal embedded between two superconducting banks (on left and right sides of the image). The overlap of the superconducting banks (dark grey) with the normal metal layer (light grey) is visible. During the measurement, the S-I-N-I-S junction is biased at a fixed current $I_{SINIS}$ and the voltage drop $V_{SINIS}$ is monitored. Bottom: Temperature dependence of the voltage $V_{SINIS}$ of the same sample with a current bias $I_{SINIS}$ = 6 pA (no current is flowing through the S-N-S junction)and zoom of the sample image.}
\label{Schema}
\end{center}
\end{figure}

Figure \ref{Schema} shows a SEM image of one of the three samples we have measured. It consists of a S-N-S junction with two additional superconducting tunnel probes on the normal metal part of the junction, thus forming a S-I-N-I-S junction (I stands for Insulator) sharing its normal metal part with the S-N-S junction. The sample was made by three-angle deposition through a suspended resist mask. Every structure is then tripled, so that the Josephson junction leads are actually made of the overlap of two superconducting (Al) layers and one normal metal (Cu) layer. The two tunnel probes were realized by depositing Al as the first layer and oxidizing it in a pressure of 5 mbar for 5 minutes prior to Cu deposition. The tunnel junction resistances are then in the range of 30-100 k$\Omega$ each. A thick (70 nm) Al deposition was finally performed in order to obtain the superconducting electrodes of the S-N-S junction. Based on our measurements, the residual resistance of the N-S contacts is estimated to be below 0.5 $\Omega$. The parameters of the samples are listed in Table 1. From the measured normal-state resistances, the diffusion constant of Cu is inferred to be about 110 cm$^2$/s for every sample. The normal metal (Cu) strip width and thickness are about 100-150 nm and 27 nm respectively. The junction length is 1, 1.5 and 2 $\mu$m for sample 1, 2 and 3 respectively. Here, the junction length $L$ is much larger than the superconducting coherence length $\xi_s$, so that the Thouless energy $E_{Th}=\hbar D / L^2$ \cite{Superlattices-Courtois} is much smaller than the superconducting gap $\Delta$. In this so-called long junction limit, the Thouless energy defines the magnitude of the proximity effects \cite{Superlattices-Courtois}, including the proximity-induced mini-gap width \cite{JLTP-Charlat,PRL-Gueron} and the critical current \cite{PRB-Dubos}. 

\begin{table}
\begin{tabular}{ccccccccc}
$\#$  & $L$ ($\mu$m) & $R_n$ ($\Omega$) & $w$ (nm) & $E_{Th}$ ($\mu$eV) & $E_{Th,fit}$ ($\mu$eV) & $\alpha$ \\
\hline
1 & 1.0 & 6.32 & 100 & 4.8 & 3.9 & 0.65\\
2 & 1.5 & 9.96 & 100 & 2.4 & 2.1 & 0.48\\
3 & 2.0  & 8.46 & 150 & 1.4 & 2.6 & 0.040
\end{tabular}
\caption{Sample parameters. The normal state resistance $R_n$, the distance between the two superconducting electrodes of the S-N-S junction $L$, the N wire width $w$ and the estimated Thouless energy $E_{Th}$ taking into account the junction length plus the overlaps of 0.1 $\mu$m with the superconducting electrodes are listed. Fit parameters of Fig. \ref{Currents} are the Thouless energy $E_{Th,fit}$ and the reduction parameter $\alpha$. The superconducting gap $\Delta$ is about 200 $\mu$eV for each sample.}
\label{samples}
\end{table}

Tunneling through a S-I-N-I-S junction is sensitive to the electron temperature in the normal metal \cite{RMP-Giazotto}. Here, we bias the S-I-N-I-S junction with a battery-powered current source in the pA range and  measure the voltage, see Fig. \ref{Schema} top. Figure \ref{Schema} bottom left shows the temperature dependence of the sample 1 S-I-N-I-S voltage at a fixed bias current of 6 pA, with zero current applied through the S-N-S junction. We obtain the expected almost linear behavior in a wide temperature range, without saturation down to below 50 mK. With this calibration, the S-I-N-I-S junction serves as an electron thermometer. In the following, we will assume an electron population in the normal metal close to quasi-equilibrium, so that we can define an effective electron temperature $T_e$, possibly different from the cryostat temperature $T_{bath}$. 

\begin{figure}[t]
\begin{center}
\includegraphics[width=8.5 cm]{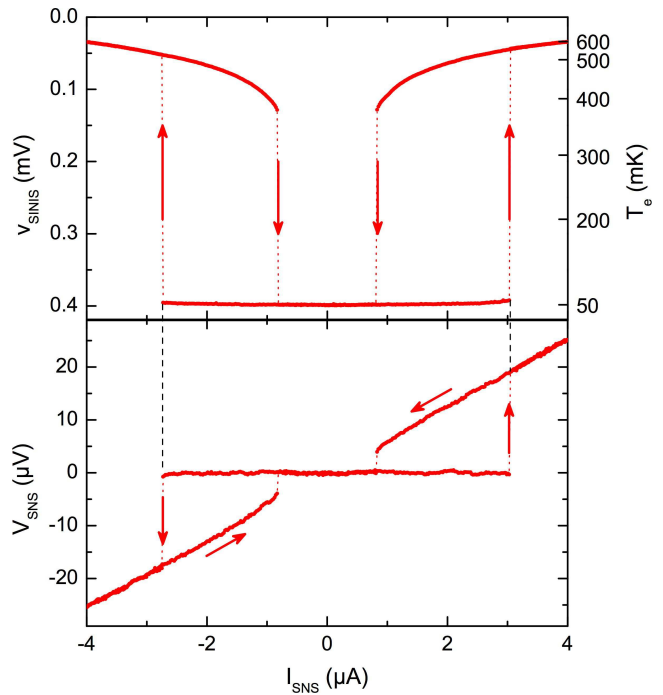}
\caption{(Color online) Current-voltage characteristic of sample 1 S-N-S junction (bottom) shown on the same current scale with the S-I-N-I-S thermometer voltage response (top) measured simultaneously at a 50 mK cryostat temperature. In the top part, the right vertical axis gives the corresponding electron temperature.}
\label{Hysteresis}
\end{center}
\end{figure}

Figure \ref{Hysteresis}, bottom part, displays the S-N-S junction current-voltage characteristic of sample 1 at the cryostat base temperature ($T_{bath}$ = 50 mK). It features a clear superconducting branch at zero voltage. When the current is increased, a sudden switch to a resistive branch with a constant resistance can be seen. As previously discussed, the characteristic is hysteretic. When the current is decreased, the voltage jumps back to zero only at a retrapping current significantly smaller than the switching current. Like the switching, the retrapping appears as a discontinuity of the characteristic. Figure \ref{Hysteresis} top part displays the simultaneously measured voltage response of the electron thermometer at a 20 pA current bias. Here we have subtracted the voltage drop in the normal metal, between the two tunnel junctions, due to the S-N-S bias current. The corresponding electron temperature $T_e$ scale is given. A striking behavior is observed. In the supercurrent branch, the electron temperature is almost constant as expected. Yet it changes slightly because of heating by the current through resistive filters on the sample stage. At the switch to the resistive branch, the electron temperature jumps to a much higher value. After the jump, the electron temperature still increases because of the increased Joule power. When the current is decreased, the electron thermometer signal first follows the same curve. Temperature stays elevated until it drops to the bath temperature value, precisely at the retrapping current. This demonstrates clearly that the hysteresis in our S-N-S junctions is governed by over-heating of the normal metal. The same measurement was performed on the same sample 1 at a different current bias of 12 pA, giving the same electronic temperature evolution, and on samples 2 and 3, displaying similar behavior.

\begin{figure}[t]
\begin{center}
\includegraphics[width=8.5 cm]{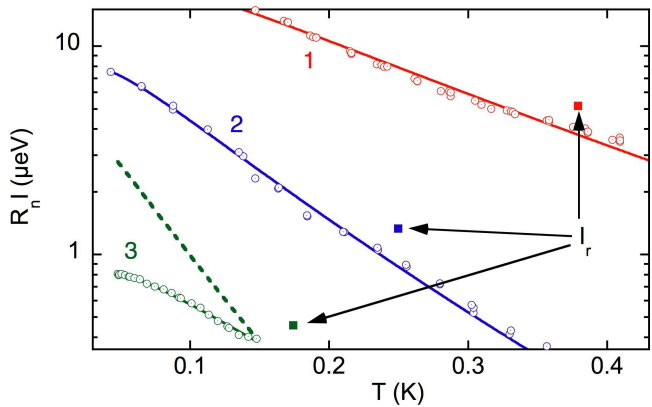}
\caption{(Color online) Temperature dependence of the $R_nI_c$ product, where $I_c$ is the measured critical current, for samples 1-3 (open symbols). The fits are displayed as full lines. For sample 3, the theoretical prediction using $\alpha$ = 0.5 and $E_{Th}$ = 1.12 $\mu$eV (close to the expected value) is shown as a dotted line. The $R_nI_r$ product including the retrapping current $I_r$ at a bath temperature of 50 mK is plotted for each sample versus the electron temperature just before retrapping (full symbols, indicated by arrows).}
\label{Currents}
\end{center}
\end{figure}

Before turning to a quantitative analysis of our experimental data, let us discuss the possible out-of-equilibrium effects. This will be done by comparing the retrapping and the switching current data. The measured temperature dependence of the $R_n I_c$ product for each sample is plotted in Fig. \ref{Currents}. We fitted the data to the theoretical prediction \cite{PRB-Dubos} using as fit parameters the Thouless energy and a scaling parameter $\alpha$ accounting for the non-ideality of the N-S interfaces. For the two shorter samples (1, 2), we obtain a good fit with values of the Thouless energy close to the estimates and values for $\alpha$ of about 0.5 \cite{PRL-Lefloch}. Although fabricated on the same chip as sample 2, sample 3 shows a different behavior with a reduced critical current and an increased effective Thouless energy, which we interpret as due to the finite phase coherence length compared to the junction length. In Fig. \ref{Currents}, we also plot the $R_nI_r$ product versus the electronic temperature before retrapping (square symbols). Here $I_r$ is the retrapping current at a bath temperature of 50 mK. In a quasi-equilibrium hypothesis, retrapping occurs when the bias current is equal to the critical current at the electron temperature so that the latter points should coincide with the equilibrium critical current data. The retrapping data are actually close to the equilibrium data with a shift of about 50 mK or less than 50 $\%$ in critical current amplitude. This limited discrepancy shows that a discussion in terms of effective electron temperature is reasonably justified.

\begin{figure}[t]
\begin{center}
\includegraphics[width=8.5 cm]{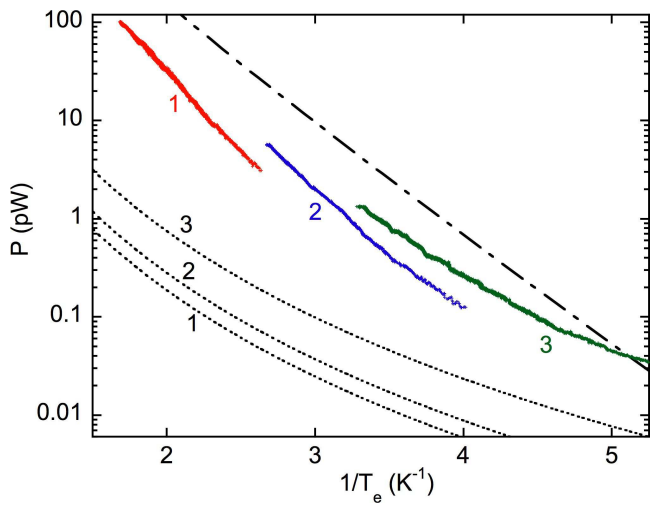}
\caption{(Color online) Injected Joule power $I_{SNS}V_{SNS}$ in the S-N-S junction as a function of the inverse of the measured electronic temperature $T_e$. The cryostat temperature is 50 mK. The expected behavior due to the electron-phonon coupling is plotted as dotted lines. The dash-dotted line is the power calculated using Eq. \ref{PowerExpr} and $\Delta$ = 200 $\mu$eV. The numbers refer to the samples.}
\label{Power}
\end{center}
\end{figure}

An elementary idea to analyze the electron heating is to consider the hot electrons to be confined in the normal metal. In this case, the electron-phonon interaction ensures the coupling to the thermal bath \cite{PRB-Wellstood}. In the limit of a low temperature for the phonons, the power flow writes $P=\Sigma U T_e^5$ with $\Sigma$ = 2 nW.$\mu$m$^{-3}$.K$^{-5}$ in Cu and $U$ is the metal volume. Here, this power flow is equal to the Joule power $I_{SNS}V_{SNS}$ dissipated in the S-N-S junction. In the case of sample 2 at 1 $\mu$A injected current, the predicted quasi-equilibrium temperature $T_e = (P/\Sigma U)^{1/5}$ is about 1 K. Figure \ref{Power} displays, as dotted lines, the power calculated within this hypothesis for the three different samples, in parallel with the experimental data, as a function of the inverse of the measured electronic temperature $T_e$. At a given temperature, the experimental data is well above the prediction, which means that electrons thermalize via another process.

We consider now the thermal link through the superconducting electrodes attached to the normal metal. The two tunnel N-I-S junctions are expected to be good thermal insulators compared to the transparent N-S interfaces. In the superconducting state, quasi-particles with an energy above the superconducting gap contribute to the heat transport through the interface and the superconductor. This was recently discussed in the framework of noise measurements in a S-N-S junction in the hot electron regime \cite{EPJ-Hoffmann}. In the following, we neglect the thermal resistance of the S-N interface compared to that of the superconducting line. The ratio $r$ between the thermal conductivities in the superconducting state and in the normal state writes \cite{Abrikosov}
\begin{equation}
r(T) =\frac{3}{2 \pi^2}\int^{+\infty}_{\Delta/k_B T}(\frac{x}{\cosh(x/2)})^2dx.
\label{ThermalCond}
\end{equation}
The normal-state thermal conductivity is $g_N^T=L_0 g_N T$, with $g_N$ being the normal state electrical conductivity. Here, we assume the Wiedemann-Franz law with the Lorentz number $L_0$=2.45.10$^{-8}$ W.$\Omega$.K$^{-2}$. In the low temperature limit, the quantity $r(T)$ decays exponentially with the temperature as $\exp(-\Delta/k_B T_e)$. 

The power flow as a function of the electronic temperature can then be calculated as an integral of the temperature-dependent thermal conductivity:
\begin{equation}
P(T_e)=L_0 G_N^{sc} \int^{T_e}_{T_{bath}} r(T) T dT
\label{PowerExpr}
\end{equation}
where $G_N^{sc}$ is the normal state electrical conductance of the superconductor from the hot region to the thermal bath. Here, we expect the Cu film to be well thermalized and act as a thermal bath. The relevant conductance is thus determined by the 70 nm-thick Al electrode between the N-S interface and the overlap with the Cu film. This corresponds to a resistance of about 9 square resistances on both electrodes (see Fig. \ref{Schema}). Assuming a typical resistivity of 2 $\mu\Omega$.cm for Al, this gives $G_N^{sc} \simeq$ 1.3 $\Omega$. The related normal-state thermal conductance $L_0 G_N^{sc}$ is then about 19 nW/K at 1 K.

The power calculated with this parameter is displayed as a function of the inverse of the electronic temperature $T_e$ in Fig. \ref{Power} (dash-dotted line). We took for the S-N-S junction a superconducting gap value $\Delta$ = 200 $\mu$eV equal to the one measured in the S-I-N-I-S junction. The calculation compares favorably with the experimental data. On a semi-logarithmic plot, both show a nearly linear decay of the power with the temperature inverse. In most of the investigated range, the measured power is lower than the prediction, which means that the thermal conductance of the superconductor was over-estimated. Nevertheless, the fair agreement between the data and our simple model shows that the electron temperature increase is actually limited by the thermal conductance of the S-N-S junction superconducting electrodes.

In conclusion, our study solves a long-standing issue in the general field of Josephson junctions by showing that the hysteresis routinely observed in long S-N-S junctions is of thermal origin. This interpretation should definitely hold also in shorter metal-based S-N-S junctions where the dissipated power density at the switching is larger. In the case of Josephson junctions based on nanowires \cite{PRL-Bezryadin,Science-DeFranceschi}, 2-DEGs \cite{PRB-Krasnov}, carbon nanotubes \cite{NatureNanotech-WW} or graphene \cite{Nature-Morpurgo}, the power density at the switching is estimated to be about or above 1 nW/$\mu$m$^3$, while the samples investigated here feature generally a smaller density in the range 2.10$^{-3}$ to 1 nW/$\mu$m$^3$. This suggests that the observed hysteresis in these other kinds of lateral Josephson junctions is also due to electron heating.

We acknowledge the financial support from the ANR contract "Elec-EPR", the ULTI-3 and NanoSciERA "Nanofridge" EU projects. We thank F. Lefloch, T.T. Heikkil\"{a} and N. Kopnin for useful discussions.


\vspace{-0.7 cm}
\begin{thebibliography}{1}
\bibitem{Likharev} K. K. Likharev, {\it Dynamics of Josephson Junctions and Circuits} (Gordon and Breach, 1991).
\bibitem{JAP-Fulton} T. A. Fulton and L. N. Dunkleberger, {\it J. Appl. Phys.} {\bf 45}, 2283 (1973).
\bibitem{JAP-Skocpol} W. J. Skocpol, M. R. Beasley and M. Tinkham, {\it J. Appl. Phys.} {\bf 45}, 4054 (1974).
\bibitem{JAP-Song} Y. Song, {\it J. Appl. Phys.} {\bf 47}, 2651 (1975).
\bibitem{PRL-Bezryadin} A. Rogachev, A. T. Bollinger and A. Bezryadin, {\it Phys. Rev. Lett.} {\bf 94}, 017004 (2005).
\bibitem{PRB-Angers}  L. Angers, F. Chiodi, G. Montambaux, M. Ferrier, S. Gu\'eron, H. Bouchiat, and J. C. Cuevas, {\it Phys. Rev. B} {\bf 77}, 165408 (2008).
\bibitem{PRB-Birge} M. S. Crosser, Jian Huang, F. Pierre, P. Virtanen, T. T. Heikkil\"{a}, F. K. Wilhelm, and N. O. Birge, {\it Phys. Rev. B} {\bf 77}, 014528 (2008).
\bibitem{PRB-Krasnov} V. M. Krasnov, T. Golod, T. Bauch and P. Delsing, {\it Phys. Rev. B} {\bf 76}, 224517 (2007).
\bibitem{Science-DeFranceschi} Y.-J. Doh, J. A. van Dam, A. L. Roest, E. P. A. M. Bakkers, L. P. Kouwenhoven and S. de Franceschi, {\it Science} {\bf 309}, 272 (2005).
\bibitem{NatureNanotech-WW} J.-P. Cleuziou, W. Wernsdorfer, V. Bouchiat, T. Ondarcuhu and M. Monthioux, {\it Nature Nanotech.}Ê{\bf 1}, 53 (2006).
\bibitem{Nature-Morpurgo} H. B. Heersche, P. Jarillo-Herrero, J. B. Oostinga, L. M. K. Vandersypen, and A. F. Morpurgo, {\it Nature} {\bf 446}, 56 (2007).
\bibitem{PRB-Tinkham} M. Tinkham, J.U. Free, C.N. Lau, and N. Markovic, {\it Phys. Rev. B} {\bf 68}, 134515 (2003).
\bibitem{RMP-Giazotto} F. Giazotto, T. T. Heikkil\"a, A. Luukanen, A. M. Savin and J. P. Pekola,  {\it Rev. Mod. Phys.} {\bf 78}, 217 (2006).
\bibitem{Superlattices-Courtois} H. Courtois, P. Gandit, B. Pannetier and D. Mailly, {\it Superlatt. and Microstruct.} {\bf 25}, 721 (1999).
\bibitem{JLTP-Charlat} F. Zhou, P. Charlat, B. Spivak and B. Pannetier, {\it J. Low Temp. Phys.} {\bf 110}, 841 (1998).  
\bibitem{PRL-Gueron} S. Gu\'eron, H. Pothier, N. O. Birge, D. Est\`eve, and M. H. Devoret, {\it Phys. Rev. Lett.} {\bf 77}, 3025 (1996).
\bibitem{PRB-Dubos} P. Dubos, H. Courtois, B. Pannetier, F.K. Wilhelm, A.D. Zaikin and G. Sch\" on, {\it Phys. Rev. B}  {\bf 63}, 064502 (2001).
\bibitem{PRL-Lefloch} E. Lhotel, O. Coupiac, F. Lefloch, H. Courtois and M. Sanquer,  {\it Phys. Rev. Lett.} {\bf 99}, 117002 (2007).
\bibitem{PRB-Wellstood} F. C. Wellstood, C. Urbina and J. Clarke, {\it Phys. Rev. B} {\bf 49}, 5942 (1994).
\bibitem{EPJ-Hoffmann} C. Hoffmann, F. Lefloch, and M. Sanquer, {\it Eur. Phys. J. B} {\bf 29}, 629 (2002).
\bibitem{Abrikosov} A. A. Abrikosov, {\it Fundamentals of the Theory of Metals}, (Elsevier Science, Amsterdam, 1988).
\end{thebibliography}
 \end{document}